\documentclass[12pt]{article}
\usepackage{graphicx}
\begin{document}

{\bf Phase Transitions on Fractals and Networks}

\bigskip

D. Stauffer

\bigskip
Institute for Theoretical Physics, Cologne University, 

D-50923 K\"oln, Euroland

\bigskip
{\bf \large Article Outline}
\bigskip

Glossary

I. Definition and Introduction

II. Ising Model

III. Fractals

IV. Diffusion on Fractals

V. Ising Model on Fractals

VI. Other Subjects ?

VII. Networks

VIII. Future Directions

\bigskip
{\bf \large Glossary}
\bigskip

{\bf Cluster}

Clusters are sets of occupied neighbouring sites.

\medskip
{\bf Critical exponent}

At a critical point or second-order phase transition, many quantities diverge
or vanish with a power law of the distance from this critical point; the 
critical exponent is the exponent for this power law.

\medskip
{\bf Diffusion}

A random walker decides at each time step randomly in which direction to 
proceed. The resulting mean square distance normally is linear in time.

\medskip
{\bf Fractals}

Fractals have a mass varying with some power of their linear dimension. The
exponent of this power law is called the fractal dimension and is smaller
than the dimension of the space.

\medskip
{\bf Ising model} 

Each site carries a magnetic dipole which points up or down; neighbouring
dipoles ``want'' to be parallel.

\medskip
{\bf Percolation} 

Each site of a large lattice is randomly occupied or empty.

\bigskip
\section{Definition and Introduction}

Some phase transitions, like the ferromagnetic Curie point where the 
spontaneous magnetisation vanishes, happen in solids, and experiments often
try to grow crystals very carefully such that the solid in which the transition
will be observed is periodic with very few lattice faults. Other phase 
transitions like the boiling of water or the liquid-vapour critical point, 
where the density difference between a liquid and its vapour vanishes, happen 
in a continuum without any underlying lattice structure. Nevertheless, 
the critical exponents of the Ising model on a simple-cubic lattice agree well
with those of liquid-vapour experiments. Impurities, which are either fixed
(``quenched dilution'') or mobile (``annealed dilution''), are known to change
these exponents somewhat, e.g. by a factor $1-\alpha$, if the specific heat 
diverges in the undiluted case at the critical point, i.e. if the specific 
heat exponent $\alpha$ is positive. In this review we deal neither with 
regular lattices nor with continuous geometry, 
but with phase transitions on fractal and other networks. We will compare these 
results with the corresponding phase transitions on infinite periodic lattices
like the Ising model.

\section{Ising Model}

Ernst Ising in 1925 (then pronounced EEsing, not EYEsing) published a model 
which is, besides percolation, one of the simplest models for phase transitions.
Each site $i$ is occupied by a variable $S_i = \pm 1$ which physicists often 
call a spin but which may also be interpreted as a trading activity \cite{cont}
on stock markets, as a ``race'' or other ethnic group in the formation of city 
ghettos \cite{schelling}, as the type of molecule in binary fluid mixtures
like isobutyric acid and water, as occupied or empty in a lattice-gas model
of liquid-vapour critical points, as an opinion for or against the government
\cite{sznajd,encynowak}, or whatever binary choice you have in mind. Also models
with more than two choices, like $S_i = -1$, 0 and 1 have been investigated 
both for atomic spins as well as for races, opinions, $\dots$. Two spins  $i$
and $k$ interact with each other by an energy $-J S_i S_k$ which is $-J$ if 
both spins are the same and $+J$ if they are the opposite of each other. Thus
$2J$ is the energy to break one bond, i.e. to transform a pair of equal spins 
to a pair of opposite spins. The total interaction energy is thus 
$$E = -J \sum_{<i,k>} S_i S_k \quad , \eqno(1a)$$
with a sum over all neighbour pairs.
If you want to impress your audience, you call this energy a Hamiltonian or 
Hamilton operator, even though most Ising model publications ignore the
difficulties of quantum mechanics except for assuming the discrete nature of 
the $S_i$. (If instead of these discrete one-dimensional values you want to
look at vectors rotating in two- or three-dimensional space, you should 
investigate the XY or Heisenberg models instead of the Ising model.) 

Different
configurations in thermal equilibrium at absolute temperature $T$ appear with
a Boltzmann probability proportional to exp$(-E/k_BT)$, and the Metropolis 
algorithm of 1953 for Monte Carlo computer simulations flips a spin with 
probability exp$(-\Delta E/k_BT)$, where $k_B$ is Boltzmann's constant and 
$\Delta E = E_{\rm after} - E_{\rm before}$ the energy difference caused by
this flip. If one starts with a random distribution of half the spins up and 
half down, using this algorithm at positive but low temperatures, one sees
growing domains. Within each domain, most of the spins are parallel, and thus 
a computer printout shows large black domains coexisting with large white 
domains. Finally, one domain covers the whole lattice, and the other spin 
orientation is restricted to small clusters or isolated single spins within that
domain. This self-organisation (biologist may call it ``emergence'') of domains
and of phase separation appears only for $0 < T < T_c$ and only in more than
one dimension. For $T > T_c$ (or at all positive temperatures in one dimension)
we see only finite domains which no longer engulf the whole lattice. This 
phase transition between long-range order below and short-range order above 
$T_c$ is called the Curie or critical point; we have $J/k_BT_c = 
{\frac{1}{2}}\ln(1 + \sqrt 2)$ on the square lattice and 0.221655 on the 
simple cubic lattice with interactions to the $z$ nearest lattice neighbours;
$z = 4$ and 6, respectively. The mean field approximation becomes valid for
large $z$ and gives $J/k_BT_c = 1/z$. Near $T = T_c$ the difference between the
number of up and down spins vanishes as $(T_c-T)^\beta$ with $\beta = 1/8$ in 
two, $\simeq 0.32$ in three, and 1/2 in six and more dimensions and in mean 
field approximation. 

We may also influence the Ising spins though an external field $h$ by adding
$$ -h \sum_i S_i    \eqno(1b) $$
to the energy of Eq.(1a). This external field then pushes the spins to become
parallel to $h$. Thus we no longer have emergence of order from the 
interactions between the spins, but imposition of order by the external field. 
In this simple version of the Ising model there is no sharp phase transition
in the presence of this field; instead the spontaneous magnetisation (fraction
of up spins minus fraction of down spins) smoothly sinks from one to zero if
the temperature rises from zero to infinity.

\section{Fractals}

Fractals obey a power law relating their mass $M$ to their radius $R$:
$$ M \propto R^D \eqno (2)$$
where $D$ is the fractal dimension. An exactly solved example are random
walks (= polymer chains without interaction) where $D = 2$ if the length of the
walk is identified with the mass $M$. For self-avoiding walks (= polymer chains
with excluded volume interaction), the Flory approximation gives $D=(d+2)/3$ 
in $d \le 4 $ dimensions ($D(d \ge 4) = 2$ as for random walks), which is exact
in one, two and four dimensions, and too small by only about two percent in 
three dimensions. 

We now discuss the fractal dimension of the Ising model. In an infinite system
at temperatures $T$ close to $T_c$, the difference $M$ between the number of 
up and down spins varies as $(T_c-T)^\beta$ while the correlation length $\xi$ 
varies as $|T-T_c|^{-\nu}$. Thus, $M \propto \xi^{-\beta/\nu}$. The 
proportionality factor varies as the system size $L^d$ in $d$ dimensions since
all spins are equivalent. In a finite system
right at the critical temperature $T_c$ we replace $\xi$ by $L$ and thus have 
$M \propto L^{d-\beta/\nu} = L^D$ with the fractal dimension 
$$ D = d - \beta/\nu \quad (d \le 4) \quad . \eqno (3a)$$

Warning: one should not apply these concepts to spin clusters if clusters are 
simply defined as sets of neighbouring parallel spins; to be fractals at $T=T_c$ 
the clusters have to be sets of neighbouring parallel spins connected by active 
bonds, where bonds are active with probability $1 - \exp(-2J/k_BT)$. Then 
the largest cluster at $T=T_c$ is a fractal with this above fractal dimension. 
 
This warning is no longer valid for percolation theory (see separate reviews in 
this encyclopedia) where each lattice site is occupied randomly with 
probability $p$ and clusters are defined as sets of neighbouring occupied sites.
For $p > p_c$ one has an infinite cluster spanning from one side of the sample
to the other; for $p < p_c$ one has no such spanning cluster; for $p=p_c$ one 
has sometimes such spanning clusters, and then the largest or spanning cluster
has 
$$M \propto L^D; \quad D = d -\beta\/\nu \quad (d \le 6) \eqno (3b)$$ 
with the critical exponents $\beta, \; \nu$ of percolation instead of Ising 
models.

These were probabilistic fractal examples, as opposed to deterministic ones like
the Sierpinski carpets and gaskets, which approximate in their fractal 
dimensions the percolation problem. We will return to them in the section
``Ising models on fractals''.

Now, instead of asking how phase transitions produce fractals we ask what 
phase transitions can be observed on these fractals.

\section{Diffusion on Fractals}

\begin{figure}[hbt]
\begin{center}
 \includegraphics[angle=-90,scale=0.5]{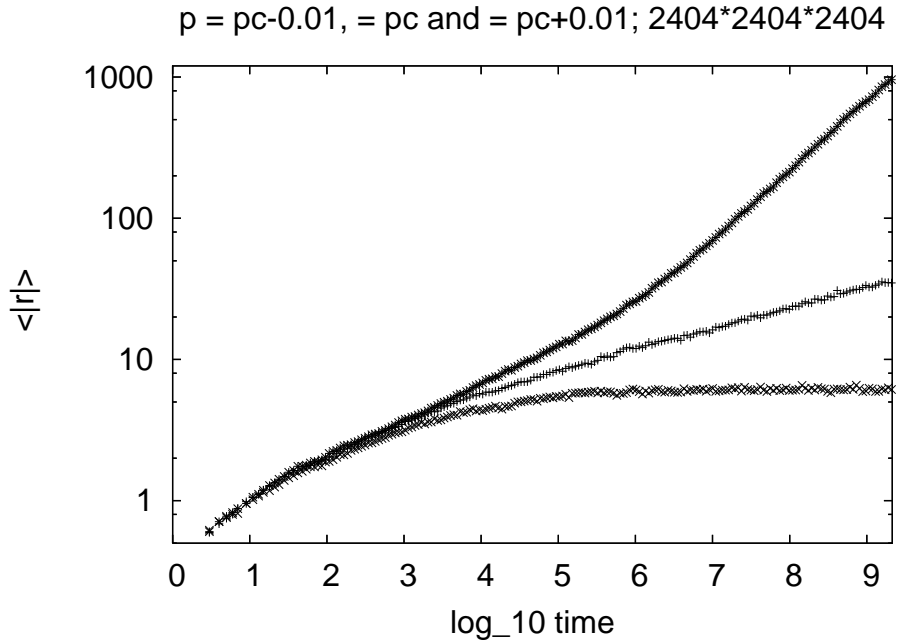}
\end{center}
\caption{Log-log plot for unbiased diffusion at (middle curve), above (upper 
data) and below (lower data) the percolation threshold $p_c$. We see the 
phase transition from limited growth at $p_c -0.01$ to diffusion at $p_c +0.01$
separated by anomalous diffusion at $p_c$. 
Average over 80 lattices with 10 walks each.
}
\end{figure}

\begin{figure}[hbt]
\begin{center}
\includegraphics[angle=-90,scale=0.5]{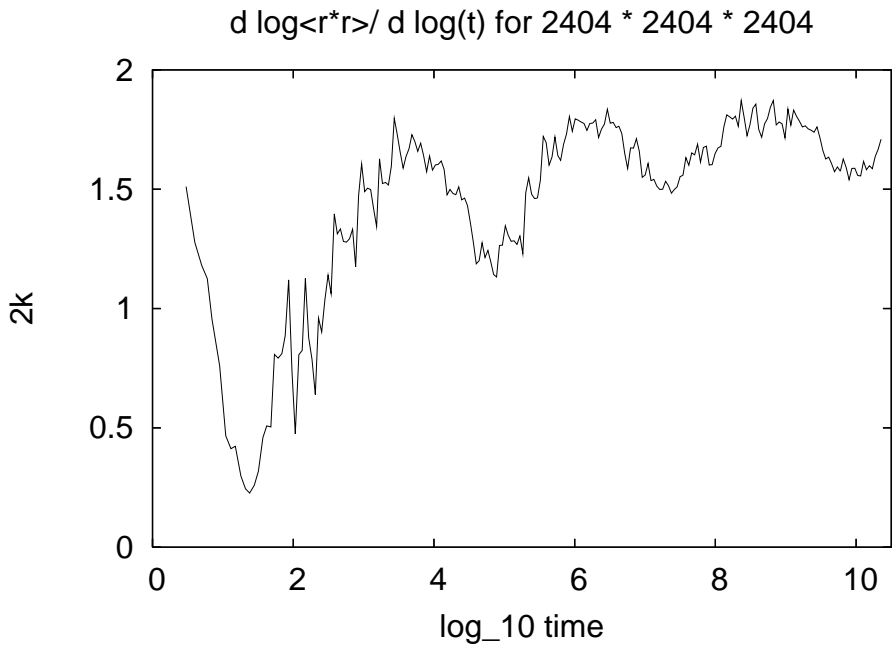}
\end{center}
\caption{Log-periodic oscillation in the effective exponent $k$ for biased 
diffusion; $p=0.725, \; B=0.98$. The limit $k=1$ corresponds to drift. 80
lattices with 10 walks each.
}
\end{figure}

\begin{figure}[hbt]
\begin{center}
\includegraphics[angle=-90,scale=0.5]{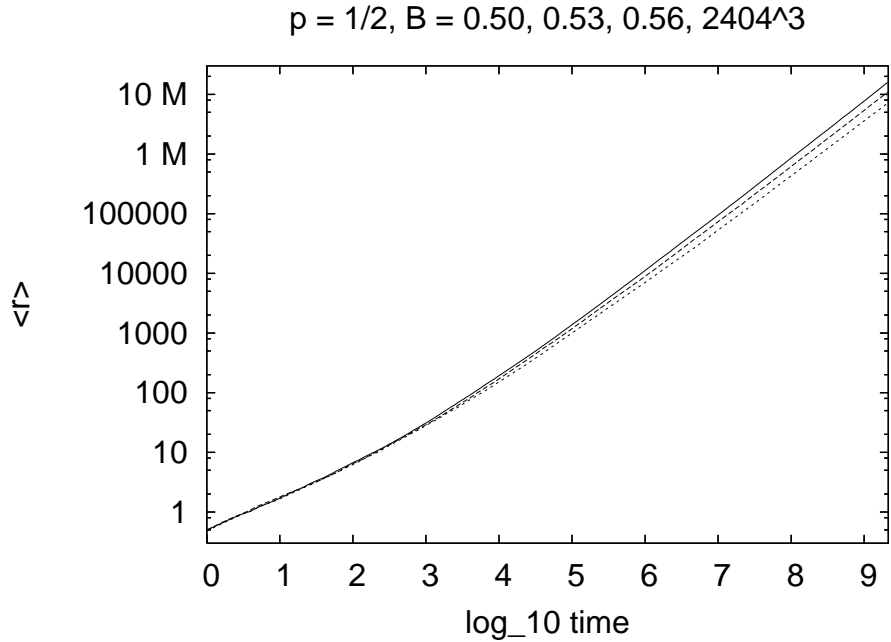}
\end{center}
\caption{Difficulties at transition from drift (small bias, upper data) to 
slower motion (large bias, lower data); 80 lattices with 10 walks each.
}
\end{figure}

\begin{figure}[hbt]
\begin{center}
\includegraphics[angle=-90,scale=0.5]{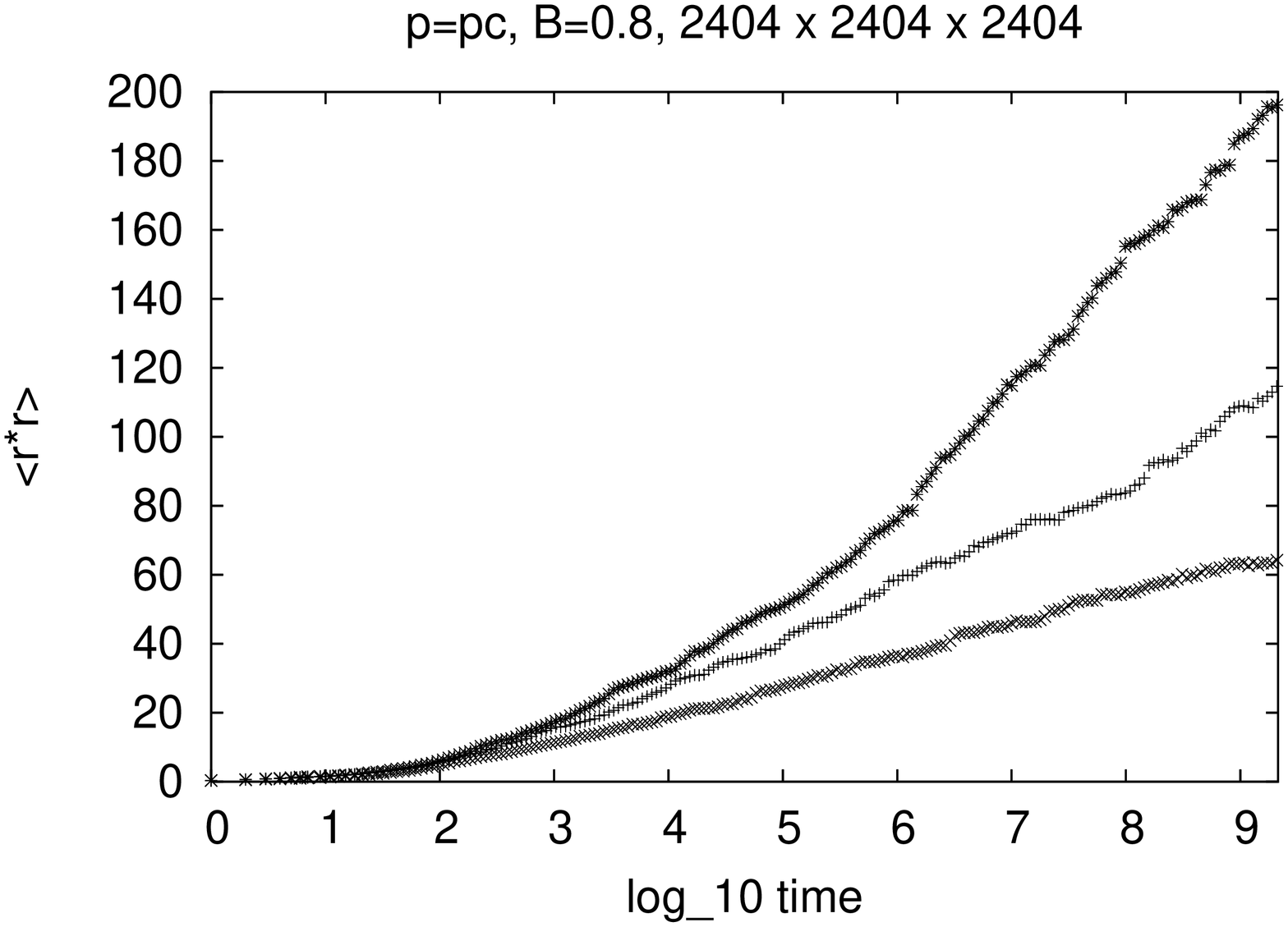}
\end{center}
\caption{Biased diffusion at $p=pc$ (middle curve) and $p = p_c \pm 0.01$ 
(upper and lower data) for bias $B = 0.8$; 80 lattices with 10 walks each.
}
\end{figure}

\subsection{Unbiased diffusion}

The most thoroughly investigated phase transitions on fractals are presumably
random walkers on percolation clusters \cite{benavraham}, particularly at 
$p = p_c$. This research was started by Brandt \cite{brandt} but it was 
the later Nobel laureate de Gennes \cite{degennes} who gave it the catchy name 
``ant in the labyrinth''. The anomalous diffusion \cite{havlin,gefen,kehr} then
made it famous a few years later and may have also biological applications
\cite{frey}.

We put an ant onto a randomly selected occupied site in the middle of a 
large lattice, where each site is permanently occupied (randomly with 
probability $p$) or empty ($1-p$). At each time step, the ant selects randomly
a neighbour direction and moves one lattice unit in this direction if and only
if that neighbour site is occupied. We measure the  mean distance
$$ R(t) = <r(t)^2>^{1/2} \quad {\rm or} \quad = <r(t)> \eqno (4) $$  
where ${\bf r}$ is the vector from the starting point of the walk and the 
present position, and $r = |{\bf r}|$ its length. The average $<\dots>$ goes 
over many such walking ants and disordered lattices. These ants are
blind, that means they do not see from their old place whether or not the 
selected neighbour site is accessible (occupied) or prohibited (empty).
(Also myopic ants were grown which select randomly always an occupied neighbour
since they can see over a distance of one lattice unit.) The squared distance
$r^2$ is measured by counting how often the ant moved to the left, to the 
right, to the top, to the bottom, to the front, or to the back on a simple
cubic lattice. 

The problem is simple enough to be given to students as a programming project.
They then should find out by their simulations that for $p < p_c$ the above
$R$ remains finite while for $p > p_c$ is goes to infinity as $\sqrt t$, for
sufficiently long times $t$. But even for $p >p_c$ it may happen that for 
a single ant the distance remains finite: If the starting point happened to 
fall on a finite cluster, then $R(t \to \infty)$ measures the radius of that
cluster. Right at $p = p_c$, instead of a constant or a square-root law, we
have anomalous diffusion: 
$$ R \propto t^k, \quad k=(\nu-\beta/2)/(2\nu+\mu-\beta) \eqno (5) $$  
for sufficiently long times. This exponent $k$ is close to but not exactly
 1/3 in two and 1/5 in three dimensions. $\beta$ and $\nu$ are the already 
mentioned percolation exponents, and $\mu$ is the exponent for the conductivity 
if percolation is interpreted as a mixture of electrically conducting and 
insulating sites. If we always start the ant walk on 
the largest cluster at $p=p_c$ instead of on any cluster, the formula for the
exponent $k$ simplifies to $\nu/(2\nu+\mu-\beta)$. The theory is explained in
detail in the standard books and reviews \cite{books,benavraham}. We see here
how the percolative phase transition influences the random walk and introduces 
there a transition between diffusion for $p > p_c$ and finite motion for 
$p < p_c$, with the intermediate ``anomalous'' diffusion (exponent below 1/2) 
at $p = p_c$. Fig.1 shows this transition on a large cubic lattice.

\subsection{Biased diffusion}
Another type of transition is seen in biased diffusion, also for $p > p_c$. 
Instead of selecting all neighbours randomly, we do that only with probability 
$1-B$, while with probability $B$ the ant tries to move in the positive 
$x$-direction. One may think of an electron moving through a disordered lattice
in an external electric field. For a long time experts discussed whether for 
$p > p_c$ one has 
a drift behaviour (distance proportional to time) for small $B$,  and a slower
motion for larger $B$, with a sharp transition at some $p$-dependent $B_c$. 
In the drift regime one may see log-periodic oscillations $\propto
\sin({\rm const} \log t)$ in the approach towards the long-time limit, Fig.2.
Such oscillations have been predicted for stock markets \cite{nikkei}, where
they could have made us rich, but for diffusion they hamper the analysis.
They come presumably from sections of occupied sites which allow motion in the
biased direction and then end in prohibited sites \cite{kirsch}.

Even in a region without such oscillations, Fig.3 shows no clear transition 
from drift to no drift; that transition could only be seen by a more 
sophisticated analysis which showed for the $p$ of Fig.3 that the reciprocal 
velocity, plotted versus log(time), switches from concave to convex shape 
at $B_c \simeq 0.53$. Fortunately, only a few years after these simulations
\cite{dhar} the transition was shown to exist mathematically \cite{ganten}.

These simulations were made for $p > p_c$; at $p=p_c$ with a fractal largest
cluster, drift seems impossible, and for a fixed $B$ the distance varies 
logarithmically, with a stronger increase slightly above $p_c$ and a limited 
distance slightly below $pc$, Fig.4. 

\section{Ising Model on Fractals}

What happens if we set Ising spins onto the sites of a fractal? In particular,
but also more generally, what happens to Ising spins on the occupied site of
a percolation lattice, when each site is randomly occupied with probability $p$?
In this case one expects three sets of critical exponents describing how the
various quantities diverge or vanish at the Curie temperature $T_c(p)$. For
$p=1$ one has the standard Ising model with the standard exponents. If $p_c$ is
the percolation threshold where an infinite cluster of occupied sites starts 
to exist, then one has a second set of exponents for $p_c < p < 1$, where 
$0 < T_c(p) < T_c(p=1)$. Finally, for zero temperature as a function of $p-p_c$
one has the percolation exponents as a third set of critical exponents. (If
$p = p_c$ and the temperature approaches $T_c(p_c) = 0$ from above, then
instead of powers of $T-T_c$ exponential behaviour is expected.) In computer
simulations, the second set of critical exponents is difficult to observe;
due to limited accuracy the effective exponents have a tendency to vary 
continuously with $p$. 

The behaviour at zero temperature is in principle trivial: each cluster of
occupied neighbor has parallel spins, the spontaneous magnetisation is given 
by the largest cluster while the many finite cluster cancel each other in their
magnetisation. However, the existence of several infinite clusters at $p=p_c$ 
disturbs this argument there; presumably the total magnetisation (i.e. 
not normalised at magnetisation per spin) is a fractal with the fractal 
dimension of percolation theory. 

Deterministic fractals, instead of the random ``incipient infinite cluster''
at the percolation threshold, may have a positive $T_c$ and then allow 
a more usual study of critical exponents at that phase transition. Koch curves
and Sierpinski structures have been intensely studied in that aspect since 
decades. To build a Sierpinski carpet we take a square, divide each side into three 
thirds such that the whole square is divided into nine smaller squares, and 
then we take away the central square. On each of the remaining eight smaller 
squares this procedure is repeated, diving each into nine squarelets and 
omitting the central squarelet. This procedure is repeated again and again, 
mathematically ad infinitum. Physicists like more to think in atoms of 
a fixed distance and would rather imagine each square to be enlarged in each 
direction by a factor three with the central square omitted; and then again 
and again this enlargement is repeated. In this way we grow a large structure
built by squares of unit area. 

Unfortunately, the phase transitions on these fractals depend on details and 
are not already fixed if the fractal dimension is fixed. Also other properties 
of the fractals like their ``ramification'' are important \cite{mandel}; see
\cite{bab} for recent work. This is highly regrettable since modern statistical
physics is not restricted to three dimensions. Models were studied also in
seven and in minus two dimensions, in the limits of dimensionality going to 
infinity or to zero,
or for non-integral dimensionality. (Similarly, numbers were generalized from 
positive counts to negative integers, to rational and irrational numbers, 
and finally to imaginary/complex numbers.) It would have been nice if these 
fractals would been models for these non-integral dimensions, giving one and 
the same set of critical exponents once their fractal dimension is known. 
Regrettably, we had to give up that hope.

Many other phase transitions, like those of Potts or voter models, were 
studied on such deterministic fractals, but are not reviewed here. We mention
that also percolation transitions exist on Sierpinski structures \cite{monceau}.
Also, various hierarchical lattices different from the above fractals show 
phase transitions, if Ising spin are put on them; the reader is referred
to \cite{berker,rozenfeld} for more literature. As the to our knowledge
most recent example we mention \cite{ausloos} that Ising spins were also thrown 
into the sandpiles of Per Bak, which show self-organised criticality.

\section{Other Subjects?}

\section{Networks}
\subsection{Definitions}

While fractals were a big physics fashion in the 1980s, networks are
now a major physics research field. 
Solid state physics requires nice single crystals where all atoms sit on a
periodic lattice. In fluids they are ordered only over shorter distances but 
still their forces are restricted to their neighbours. Human beings, on the 
other hand, form a regular lattice only rarely, e.g. in a fully occupied 
lecture hall. In a large crowd they behave more like a fluid. But normally
each human being may have contacts with the people in neighbouring residences, 
with other neigbours at the work place, but also via phone or internet with 
people outside the range of the human voice. Thus social interactions between
people should not be restricted to lattices, but should allow for more 
complex networks of connections. 
 
One may call Flory's percolation theory of 1941 a network, and the later random
graphs of Erd\"os and R\'enyi (where everybody is connected with everybody, 
albeit with a low probability) belong to the same ``universality class'' (same 
critical exponents) as Flory's percolation. In Kauffman's random Boolean 
network of 1969, everybody has $K$ neighbours selected randomly from the $N$ 
participants. Here we concentrate on two more recent network types, the 
small-world  \cite{strogatz} and the scale-free \cite{albert} networks of
1998 and 1999 respectively (with a precursor paper of economics Nobel laureate
Simon \cite{simon} from 1955).
 
The small-world or Watts-Strogatz networks start from a regular lattice, often
only a one-dimensional chain. Then each connection of one lattice site to
one of its nearest neighbours is replaced randomly, with probability $p$, by a 
connection to a randomly selected other site anywhere in the lattice. Thus the 
limits $p = 0$ and 1 correspond to regular lattices and roughly random graphs, 
respectively. 

In this way everybody may have exactly two types of connections, to nearest 
neighbours and to arbitrarily far away people. This unrealistic feature
of small-world networks is avoided by the scale-free networks of Barab\'asi 
and Albert \cite{albert}, defined only through topology with (normally) no
geometry involved:

We start with a small set of fully connected people. Then more people join the 
network, one after the other. Each new member selects connections to exactly 
$m$ already existing members of the network. These connections are not 
random but follow preferential attachment: The more people have selected 
a person to be connected with in the past, the higher is the probability that
this person is selected by the newcomer: the rich 
get richer, famous people attract more followers than normal people. In the 
standard Barab\'asi-Albert network, this probability is proportional to the
number of people who have selected that person. In this case, the average
number of people who have been selected by $k$ later added members varies 
as $1/k^3$. A computer program is given e.g. in \cite{newbook}.

These networks can be undirected (the more widespread version) or directed
(used less often.) For the undirected or symmetric case, the connections 
are like friendships: If A selects B as a friend, then B also has A as a 
friend. For directed networks, on the other hand, if A has selected B as 
a boss, then B does not have A as a boss, and the connection is like a one-way 
street. Up to $10^8$ nodes were simulated in directed scale-free networks. We 
will now check for phase transitions on both directed and undirected networks.
 
\subsection{Phase transitions}

The Ising model in one dimension does not have a phase transition at some 
positive critical temperature $T_c$. However, its small-world generalisation,
i.e. the replacement of small fraction of neighbour bonds by long-range bonds, 
produces a positive $T_c$ with a spontaneous magnetisation proportional to 
$(T_c-T)^\beta$, and a $\beta$ smaller than the 1/8 of two dimensions 
\cite{barrat}.
 
The Solomon network is a variant of the small-world network: Each person has 
one neighbourhood corresponding to the workplace and another neighborhood 
corresponding to the home \cite{malarz}. It was suggested and simulated by
physicists Solomon and Malarz, respectively, before Edmonds \cite{edmonds} 
criticised physicists for not having enough ``models which explicitely
include actions and effects within a physical space as well as communication
and action within a social space.'' Even in one dimension a spontaneous
magnetisation was found.

On Barab\'asi-Albert (scale-free) networks, Ising models were found 
\cite{aleks} for small $m$ and millions of spins to have a spontaneous 
magnetisation for temperatures below some critical temperature $T_c$ which
increases logarithmically with the number $N$ of spins: $k_BT_c/J \simeq
2.6 \ln(N)$ for $m=5$. 

Here we had undirected networks with symmetric couplings between spins: actio 
= --reactio, as required by Newton. Ising spins on directed networks, on the 
other hand, have no well-defined total energy, though each single spin may be
influenced as usual by its $m$ neighbour spins. If in an isolated pair of spins
$i$ and $k$ we have a directed interaction in the sense that spin $k$ tries to 
align spin $i$ into the direction of spin $k$, while $i$ has no influence on 
$k$, then we have a perpetuum mobile: Starting with the two spins antiparallel,
we first flip $i$ into the direction of $k$, which gives us an energy $J$. Then
we flip spin $k$ which does not change the energy. Then we repeat again and 
again these two spin flips, and gain an energy $J$ for each pair of flips: too
nice to be true. The violations of Newton's symmetry requirement makes this
directed network applicable to social interactions between humans, but not to
forces between particles in physics. 

On this directed Barab\'asi-Albert network, the ferromagnetic Ising spins 
gave no spontaneous magnetization, but the time after which the magnetisation
becomes zero (starting from unity) becomes very long at low temperatures,
following an Arrhenius law \cite{sumour}: time proportional to exp(const/$T$). 
Also on a directed lattice such Arrhenius behavior was seen while for directed
random graphs and for directed small-world lattices a spontaneous magnetisation 
was found \cite{sumour}. A theoretical understanding for these directed cases 
is largely lacking.
 
Better understood is the percolative phase transition on scale-free networks
(see end of Sec.2 for definition of percolation). If a fraction $1-p$ of the
connections in an undirected Barab\'asi-Albert network is cut randomly, does 
the remaining fraction $p$ keep most of the network together? It does, for 
large enough networks, since the percolation threshold $p_c$ below which 
no large connected cluster survives, goes to zero as 1/log($N$) where $N$ 
counts the number of nodes in the network. This explains why inspite of the 
unreliability of computer connections, the internet still allows most computers
to reach most other computers in the word: If one link is broken, some other
link may help even though it may be slower \cite{albert}.

\section{Future Directions}

We reviewed here a few phase transitions, and ignored many others. At present 
most interesting for future research seem to be the directed networks, since
they have been investigated with methods from computational physics even though
they are not part of usual physics, not having a global energy. A theoretical
(i.e. not numerical) understanding would help.

\end{document}